\documentclass[a4paper,12pt]{article}
\pdfoutput=1

\usepackage{amsmath,amssymb,amsfonts}
\usepackage{physics}
\allowdisplaybreaks
\usepackage{graphicx}
\usepackage[dvipsnames]{xcolor}
\usepackage{caption}
\usepackage{subcaption}
\usepackage[bottom]{footmisc}	
\usepackage{soul}
\makeatletter
\@addtoreset{equation}{section}
\renewcommand{\theequation}{\thesection.\@arabic\c@equation}
\makeatother
\usepackage{hyperref} 
    \hypersetup{colorlinks=true, linktocpage, linkcolor={blue}, citecolor={blue}, urlcolor={blue}} 
\usepackage{cite}
\usepackage{caption}

\def \be {\begin{equation}}
\def \ee {\end{equation}}
\def \bea {\begin{align}}
\def \eea {\end{align}}
\def \nn {\nonumber}

\def \rr {\raise.35ex\hbox{\small $\prime$}\kern-.17em{\mbox{\large $\imath$}}}
\def \del {\partial}
\def \dels {\partial\kern-.5em / \kern.5em}
\def \As {{A\kern-.5em / \kern.5em}}
\def \Ds {D\kern-.7em / \kern.5em}

\def \b {\beta}
\def \dag {\dagger}

\def \d {\delta}
\def \eps {\epsilon}

\def \lam {\lambda}

\def \om {\omega}
\def \Om {\Omega}

\def \th {\theta}
\def \Th {\Theta}

\def \t {\tau}

\newcommand{\hide}[1]{}

\pdfoutput=1

\begin{document}



\begin{titlepage}
\vspace*{-10mm}   
\baselineskip 10pt   
\begin{flushright}   
\begin{tabular}{r} 
TIT/HEP-697
\end{tabular}   
\end{flushright}   
\baselineskip 24pt   
\vglue 10mm

\begin{center}

\noindent
\textbf{\LARGE
A Stringy Effect on Hawking Radiation 
%
}
\vskip10mm
\baselineskip 20pt

\renewcommand{\thefootnote}{\fnsymbol{footnote}}

{\large
Pei-Ming~Ho${}^{a,b}$\,\footnote[1]{pmho@phys.ntu.edu.tw},
Yosuke~Imamura${}^{c}$\,\footnote[2]{imamura.y.ad@m.titech.ac.jp},
Hikaru~Kawai${}^{a,b}$\,\footnote[3]{hikarukawai@phys.ntu.edu.tw},
Wei-Hsiang Shao${}^{a,b}$\,\footnote[4]{whsshao@gmail.com}
}

\renewcommand{\thefootnote}{\arabic{footnote}}


{\it
${}^{a}$
Department of Physics and Center for Theoretical Physics, National Taiwan University, \\
No. 1, Sec. 4, Roosevelt Road, Taipei 106319, Taiwan
\\
${}^{b}$
Physics Division, National Center for Theoretical Sciences, \\
No. 1, Sec. 4, Roosevelt Road, Taipei 106319, Taiwan
\\
${}^{c}$
Department of Physics, Tokyo Institute of Technology, 
Tokyo 152-8551, Japan
}

\vskip 20mm
\begin{abstract}
\normalsize

In string theories,
interactions are exponentially suppressed for trans-Planckian space-like external momenta.
We study a class of quantum field theories that exhibit this feature
modeled after Witten's bosonic open string field theory,
and discover a Lorentz-invariant UV/IR relation 
that leads to the spacetime uncertainty principle proposed by Yoneya.
Application to a dynamical black hole background suggests that 
Hawking radiation is turned off around the scrambling time.

\end{abstract}
\end{center}

\end{titlepage}

\pagestyle{plain}

\baselineskip 18pt

\setcounter{page}{1}
\setcounter{footnote}{0}
\setcounter{section}{0}

\tableofcontents



\section{Introduction}
\label{sec:Introduction}


Despite the trans-Planckian problem~\cite{tHooft:1984kcu, Jacobson:1991gr}, 
the robustness of Hawking radiation~\cite{Hawking:1974rv, Hawking:1975vcx} 
against UV modifications is widely accepted~\cite{Hambli:1995pp, Unruh:2004zk, Agullo:2009wt, Kajuri:2018myh, Gryb:2018pur}.
Otherwise, the information loss paradox~\cite{Hawking:1976ra, Mathur:2009hf}
would not have attracted so much attention.
Among the opposing opinions~\cite{Helfer:2003va, Kawai:2013mda, Kawai:2014afa, Akhmedov:2015xwa, Ho:2020cbf, Ho:2020cvn, Ho:2021sbi, Ho:2022gpg},
it was demonstrated~\cite{Ho:2020cbf,Ho:2020cvn,Ho:2021sbi,Ho:2022gpg}
that the effective theory prediction breaks down around the scrambling time\,\footnote{
The {\em scrambling time} for a black hole of Schwarzschild radius $a$
is taken to be $4a \log(a / \ell)$~\cite{Sekino:2008he}, 
where $\ell$ is the UV length scale.
}
due to higher-derivative, non-renormalizable interactions
between the radiation field and the curved background or the collapsing matter,
as certain Lorentz-invariants become trans-Planckian.
It was also noted that the nice-slice argument~\cite{Polchinski:1995ta} 
does not exclude UV effects on Hawking radiation~\cite{Ho:2021sbi}.
The evolution of virtual particles depends on UV physics 
whenever trans-Planckian Lorentz invariants are present, 
even in the absence of high-energy events.\footnote{
For instance, 
we would need a UV theory in order to describe 
a trans-Planckian measurement in the Minkowski vacuum.
}

The claim that Hawking radiation is sensitive to UV physics 
may appear to conflict with findings related to hypothetical UV models 
such as modified dispersion relations~\cite{Unruh:1994je, Brout:1995wp, Corley:1996ar, Corley:1997pr, Himemoto:1999kd, Saida:1999ap, Jacobson:1999zk, Unruh:2004zk, Barcelo:2005fc}.
However, a recent study~\cite{Akhmedov:2023gqf} has shown
that more general UV dispersions could significantly modify Hawking radiation.
In any case,
there is growing evidence suggesting that the trans-Planckian problem 
associated with Hawking radiation 
cannot be dismissed as a mere artifact of the choice of the coordinate system.
Through this work, 
we aim to understand the behavior of Hawking radiation in string theory,
which is a prominent candidate for a theory of quantum gravity.
Notable related works include Refs.~\cite{Susskind:1993aa, Dodelson:2015toa}.

A common characteristic of string theories is 
the exponential suppression of interactions beyond the Planck scale,
and this work will focus specifically on its effect.
Motivated by superstring field theories
(as quantum field theories of spacetime fields)
\cite{Sen:2015uaa, Pius:2016jsl, deLacroix:2017lif, Pius:2018crk, DeLacroix:2018arq},
we consider a class of scalar field theories with the action\footnote{
More generally,
finite derivatives of the fields $\tilde{\phi}_a$ are allowed
in the interaction terms,
and the discussions below remain essentially the same
with this generalization.
}
\be
S = \int d^D x \left\{
\frac{1}{2}
\sum_a \phi_a \left( \Box - m_a^2 \right) \phi_a
- \sum_{a_1\cdots a_n} \frac{1}{n!} \, \lam_{a_1\cdots a_n} \tilde{\phi}_{a_1} \cdots \tilde{\phi}_{a_n}
\right\}
\label{L-def-1}
\ee
in Minkowski space,
where all of the fields $\phi_a$ appear in interactions 
only in the nonlocal form\footnote{
This is of the same form as 
Witten's bosonic open string field theory~\cite{Witten:1985cc, Kostelecky:1989nt},
in which the length parameter is given by
$\ell^2 \equiv 2 \alpha' \log \left(3\sqrt{3}/4\right)$.
In superstring field theories,
the interaction vertex $e^{- \ell^2 k^2}$ in the abovementioned theory is in general
the exponential of a quadratic function of momenta.}
\be
\tilde{\phi}_a \equiv 
e^{\ell^2 \Box / 2} \phi_a \, .
\label{tildephi-def}
\ee
We have adopted the mostly plus signature $(- , + , \cdots , +)$ in our convention,
thus
$\Box \equiv
\del_{\mu} \del^{\mu}
\equiv
- \del_{T}^2 + \grad^2$.
The model~\eqref{L-def-1} mimics the trans-Planckian suppression in string field theories,
and has been utilized as a tool for studying such theories in the literature~\cite{Pius:2016jsl, Pius:2018crk, Buoninfante:2022krn}.
Henceforth,
we will refer to it as the \emph{stringy model}.
Nevertheless,
it is not known how closely (or poorly) this string-theory-inspired model 
resembles the string theory.
It is a toy model without a controlled approximation to string theory.
Furthermore,
as the stringy model~\eqref{L-def-1} is nonlocal,
its quantization is potentially problematic.
While the unitarity of this model in the path integral formulation
has already been checked in Refs.~\cite{Pius:2016jsl, Pius:2018crk, Briscese:2018oyx, Chin:2018puw, Buoninfante:2022krn},
there can still be various pathologies that will only be uncovered 
after a thorough examination.

Naively,
since $\phi_a$ satisfies $(\Box - m_a^2) \phi_a = 0$ 
at zeroth order in the perturbation theory,
we have
$\tilde{\phi}_a \simeq e^{\ell^2 m_a^2 / 2} \phi_a$
at the leading order~\cite{Cheng:2001du,Cheng:2002rz}.
If this were the case, Hawking radiation would remain 
largely unchanged compared to the low-energy effective theory~\cite{Kajuri:2018myh}.
However,
this interpretation only holds in the low-energy limit of the model~\eqref{L-def-1}, 
where the perturbative treatment is valid~\cite{Cheng:2001du,Cheng:2002rz}.

Since string theory is proposed as a theory of \emph{everything},
both the radiation field describing Hawking particles 
and the apparatus used for their detection 
should be governed by the same action~\eqref{L-def-1}.
Therefore, 
the detection of a field $\phi_a$ is actually the detection of $\tilde{\phi}_a$,
as all interactions depend exclusively on $\tilde{\phi}_a$.
For instance,
an Unruh-DeWitt detector~\cite{Unruh:1976db, DeWitt:1980hx} 
would couple to one of the fields $\phi_a$ in the form of 
\be
\hat{H}_{\text{int}}(\t) = \lam \hat{M}(\t) \hat{\tilde{\phi}}_a \bigl[ x(\t) \bigr] \, ,
\label{Unruh-DeWitt}
\ee
where $\t$ represents the proper time along the trajectory $x(\tau)$ of the detector,
$\hat{M}(\t)$ is an operator acting on the Hilbert space of the detector,
and $\hat{\tilde{\phi}}_a$ is the corresponding field operator in the Hamiltonian formulation.
Hence,
the Unruh effect~\cite{Fulling:1972md, Davies:1974th, Unruh:1976db}
as the response of a uniformly accelerating detector to the Minkowski vacuum $\ket{0}$
is determined by the Wightman function
$\langle 0 | \hat{\tilde{\phi}}_a (x) \hat{\tilde{\phi}}_a (x') | 0 \rangle$,
rather than $\langle 0 | \hat{\phi}_a (x) \hat{\phi}_a (x') | 0 \rangle$.
By the same token,
Hawking radiation is determined by the Wightman function of $\hat{\tilde{\phi}}_a$
instead of $\hat{\phi}_a$.
In fact, all physical observables directly probed by measurements
should be expressed in terms of $\hat{\tilde{\phi}}_a$.

In view of this,
it is natural to treat $\tilde{\phi}_a$ as the fundamental fields
and express everything in the action~\eqref{L-def-1} 
in terms of $\tilde{\phi}_a$ as
\be
S = \int d^D x \left\{
\frac{1}{2}
\sum_a \tilde{\phi}_a \left( \Box - m_a^2 \right) e^{- \ell^2 \Box} \, \tilde{\phi}_a
- \sum_{a_1\cdots a_n} \frac{1}{n!} \, \lam_{a_1\cdots a_n} \tilde{\phi}_{a_1} \cdots \tilde{\phi}_{a_n}
\right\} \, ,
\label{L-def-2}
\ee
where the infinite derivatives in the interaction terms
have been transferred to the kinetic terms through the field redefinition~\eqref{tildephi-def}.
This then allows us to treat the interaction terms as perturbations
when studying the effect of the infinite derivatives on the field quantization
as well as the correlation functions of $\tilde{\phi}_a$.

As in Witten's bosonic open string field theory,
the action~\eqref{L-def-2} is used in the path integral formulation,
which needs to be evaluated via analytic continuation,
typically in the Euclideanized momentum space~\cite{Pius:2016jsl, Pius:2018crk, Buoninfante:2022krn}.
To compute Hawking radiation, all we need is the Wightman function 
$\langle 0 | \hat{\tilde{\phi}}_a (x) \hat{\tilde{\phi}}_a (x') | 0 \rangle$.
However, it is unclear how to derive the Wightman function
in the path integral formulation of a nonlocal theory.
In this work,
we find that the Wightman function can be derived in the light-cone frame
by analytically continuing the string length parameter $\ell^2 \to \ell_E^2 = -i \ell^2$,
which corresponds to a complexification of 
the string worldsheet modular parameter~\cite{Witten:2013pra}.

Interestingly,
it turns out that the Wightman function is subject to 
a Lorentz-invariant UV cutoff on the light-cone momentum,
leading to the uncertainty relation
\be
\Delta U \Delta V \gtrsim 4 \ell_E^2 \, ,
\label{DUDV-0}
\ee
which aligns with the light-cone version of the 
\emph{spacetime uncertainty principle} $\Delta T \Delta X \gtrsim \ell^2$
proposed by Yoneya~\cite{Yoneya:1987gb, Yoneya:1988, Yoneya:1989ai, Yoneya:2000bt}
as a fundamental feature of string theory.
We will show that,
as a result of eq.~\eqref{DUDV-0},
the magnitude of Hawking radiation tends towards zero 
after the scrambling time $\sim 4a \log(a / \ell)$,
although the Hawking temperature remains unchanged.
This conclusion is in line with the understanding 
that a UV cutoff at the Planck scale shuts down Hawking radiation
around the scrambling time~\cite{Ho:2022gpg}.
It also resonates with a recent study~\cite{Chau:2023zxb} 
regarding the fate of Hawking radiation 
under the \emph{generalized uncertainty principle} (GUP)~\cite{Amati:1987wq, Gross:1987kza, Gross:1987ar, Amati:1988tn, Fabbrichesi:1989ps, Konishi:1989wk, Guida:1990st, Maggiore:1993rv}
\be
\Delta X \Delta K \geq \frac{1}{2} \left[ 1 + \ell^2 (\Delta K)^2 \right] \, ,
\ee
which is designed to capture a nonlocal aspect of string theory. 
This study~\cite{Chau:2023zxb} revealed that the GUP also results in 
the termination of Hawking radiation around the scrambling time. 

This paper is structured as follows.
In Sec.~\ref{sec:StringyModels}, 
we derive the Wightman function of the stringy model 
and establish the algebra obeyed by 
the creation and annihilation operators 
in the decomposition of the field operator $\hat{\tilde{\phi}}_a$. 
The emergence of a UV/IR connection is explored,
and its associated physical interpretations are also discussed.
In Sec.~\ref{sec:HawkingRadiation},
we calculate the Hawking radiation in the stringy model,
and find that its amplitude diminishes to zero around the scrambling time.
Finally, 
we present our conclusions and comment on the implications of our results
in Sec.~\ref{sec:conclusion}.


\section{Stringy Model}
\label{sec:StringyModels}


In this section,
we start by illustrating the necessity of 
employing analytic continuation in the stringy model,
and then introduce the analytic continuation of 
the string length parameter $\ell$ in the light-cone frame.
This extension enables us to derive the Wightman function, 
which showcases a UV/IR connection 
that ultimately gives rise to 
the spacetime uncertainty principle~\eqref{DUDV-0} with a minimum length.
As we will see, 
the same property also implies a weaker interaction
between the field and background fluctuations.


\subsection{The Feynman propagator}
\label{sec:FeynmanPropagator}


From now on,
we shall focus on a massless scalar field denoted simply as $\tilde{\phi}$.
Extensions of our discussion to massive scalar fields are straightforward.
In this subsection,
we explain the importance of analytic continuation
for the stringy model~\eqref{L-def-2}.
We also illustrate how the nonlocality inherent in the theory
hinders the derivation of the Wightman function.

The Feynman propagator of $\tilde{\phi}$ 
can be read off from the action~\eqref{L-def-2} as
\begin{align}
{\cal D}(k) \equiv 
- i \, \frac{e^{- \ell^2 k^2}}{k^2 - i \epsilon} \, .
\label{propagator-Mink}
\end{align}
It differs from the standard propagator of the low-energy effective theory
by an exponential factor
\be
e^{- \ell^2 k^2} = e^{\ell^2 (k^0)^2} e^{- \ell^2 |\textbf{k}|^2} \, .
\label{expk2}
\ee
The factor $e^{\ell^2 (k^0)^2}$ is unphysical, as it is exponentially large for time-like momenta, which further leads to UV divergences in integrals over the momentum space.
For this reason,
Feynman-diagram calculations are typically carried out
in the Euclideanized momentum space~\cite{Pius:2016jsl, Pius:2018crk, Buoninfante:2022krn}.

In the position space,
the Feynman propagator is formally given by
\be
G(x) = \int\frac{d^D k}{(2\pi)^D} \, {\cal D}(k) \, e^{ik\cdot x} \, ,
\label{G(X)}
\ee
where $d^D k \equiv dk^0 d^{D-1}\textbf{k}$.
Through Wick rotation $k^0 = i k_E$,
the exponential factor $e^{- \ell^2 k_E^2}$ in the propagator~\eqref{propagator-Mink}
suppresses the contributions from trans-Planckian modes,
and the position-space representation~\eqref{G(X)}
can then be evaluated to give
\be
G(x) 
=
\frac{1 - e^{- x^2 / 4 \ell^2}}{4\pi^{2} x^{2}} \, ,
\label{G4}
\ee
where $x^2 \equiv - T^2 + \textbf{X}^2$.
There is again an unphysical factor $e^{- x^2/ 4 \ell^2}$
that grows exponentially in time-like directions.
Consequently, 
whether we consider the Feynman propagator of the stringy model
as defined in eq.~\eqref{propagator-Mink} or eq.~\eqref{G4},
it cannot be directly interpreted in the Lorentzian signature.
Analytic continuation is inevitably required.

In the context of Feynman diagrams,
there is a prescription for selecting appropriate integration contours 
for loop energies in the Euclideanized momentum space
which ensures that the Cutkosky rules hold
and that unitarity is preserved~\cite{Pius:2016jsl, Pius:2018crk, Buoninfante:2022krn}.
In this sense, 
the stringy model~\eqref{L-def-2} is a good candidate 
for investigating UV physics in terms of S-matrix calculations.

Our first task is to construct the Wightman function.
It is tempting to interpret the Feynman propagator~\eqref{G(X)} as the time-ordered product
\be
G(x - x') =
\langle 0 | \mathcal{T} \bigl( \hat{\tilde{\phi}}(x) \hat{\tilde{\phi}}(x') \bigr) | 0 \rangle \equiv \Th(T - T') W(x - x') + \Th(T' - T) W(x' - x) \, ,
\label{P=W+W}
\ee
where $x = (T, \textbf{X})$, $\Theta(T)$ is the step function, and 
\be
W(x - x')
= \langle 0 | \hat{\tilde{\phi}}(x) \hat{\tilde{\phi}}(x') | 0 \rangle
\label{W=VEV}
\ee
is the Wightman function.
However, 
this naive expectation fails for the stringy model.
Take for example in two dimensions ($D = 2$) where $x = (T, X)$ and $k = (k^0, K)$. 
Upon Wick rotations 
\be 
T = i T_E \, , \qquad k^0 = i k_E \, ,
\ee 
the stringy propagator~\eqref{G(X)} can be evaluated explicitly as
\begin{align}
G(T_E, X) 
&= 
i \int_{-\infty}^{\infty} \frac{dk_E \, d K}{4\pi^2} \, 
\frac{- i e^{- \ell^2 ( k_E^2 + K^2 )}}{(k_E)^2 + K^2} \, e^{i (k_E T_E \, + \, K X)}
\nn \\
&= 
\int_{-\infty}^{\infty} \frac{dK}{2\pi} \, \frac{e^{i K X}}{4 |K|} 
\left[
e^{- \left| K \right| \, T_E} \, \mbox{erfc}\left(|K|\ell - \frac{T_E}{2\ell}\right)
+ e^{|K| \, T_E} \, \mbox{erfc}\left(|K|\ell + \frac{T_E}{2\ell}\right)
\right] \, .
\label{Gstringy-TX}
\end{align}
The complementary error function $\mbox{erfc}(z)$
can be approximated by the step function $2\Th(- z)$ for large $|z|$,
thus eq.~\eqref{Gstringy-TX} reduces to the form of the low-energy effective theory 
in the limit $\ell \rightarrow 0$.
However,
the fact that the step function $\Theta(T_E)$
is now replaced by a smooth function $\mbox{erfc} (\abs{K} \ell - T_E / 2 \ell)$
is a reflection of the nonlocality in the stringy model,
and it is incompatible with eq.~\eqref{P=W+W} at finite $\ell$.
This prevents us from a direct deduction of the Wightman function
from the Feynman propagator.


\subsection{Wightman function in the light-cone frame}
\label{sec:WightmanFunction}


Witten's bosonic open string field theory can be formulated as a local theory
only in one light-cone direction~\cite{Erler:2004hv}.
To derive the Wightman function,
we adopt the light-cone frame
and perform an analytic continuation of the string length parameter $\ell$.

Let us begin by reviewing the standard low-energy effective theory 
in the light-cone frame,
working with just two spacetime dimensions for simplicity.
The generalization to higher dimensions is straightforward.
In terms of the light-cone coordinates defined by
\be
U \equiv T - X \, ,
\qquad
V \equiv T + X \, ,
\ee
the Feynman propagator takes the form
\begin{align}
G(U, V) = 
\int_{-\infty}^{\infty} \frac{d\Om d\bar{\Om}}{4\pi^2} \, \frac{i}{2\Om\bar{\Om} + i\epsilon} \,
e^{- i\Om U - i\bar{\Om} V} \, ,
\end{align}
where $\Omega \equiv (k^0 + K) / 2$ and $\bar{\Omega} \equiv (k^0 - K) / 2$ are the light-cone momenta.
We carry out the contour integral over $\bar{\Om}$
while assuming that $\Om \neq 0$
(ignoring contributions from the ingoing modes with $\Om = 0$).
This yields the propagator for the outgoing modes:
\begin{align}
G_{\text{out}}(U, V) 
&=
\int_{0}^{\infty} \frac{d\Om}{2\pi} \, \frac{1}{2\Om} \, 
\left[
e^{- i\Om U} \Th(V) + e^{i\Om U} \Th(-V)
\right] \, .
\label{G-standard}
\end{align}
By interpreting this propagator as a time-ordered product 
analogous to eq.~\eqref{P=W+W} but adapted for the light-cone frame, i.e.
\be
G_{\text{out}}(U, V) = 
\Th(V) W_{\text{out}}(U, V) + \Th(- V) W_{\text{out}}(-U, -V) \, ,
\label{G=W+W}
\ee
we can identify the Wightman function for the outgoing modes as
\be
W_{\text{out}}(U, V)
= \int_{0}^{\infty} \frac{d\Om}{2\pi} \, \frac{1}{2\Om} \, 
e^{- i\Om U} \, .
\label{W-EFT}
\ee

We now turn to the 2D propagator in the stringy model.
Due to the unphysical feature of exponential growth
in time-like directions in the Lorentzian signature,
we anticipate that eq.~\eqref{G=W+W} holds
only through analytic continuation.
Recall the Schwinger parametrization of the stringy Feynman propagator~\eqref{propagator-Mink}:
\be
\mathcal{D}(k) = 
- i \, \frac{e^{- \ell^2 k^2}}{k^2} = - i \int_{\ell^2}^{\infty} d\alpha \, e^{- \alpha k^2} \, ,
\label{stringy-Schwinger-1}
\ee
which is similar to the expression for the ordinary Feynman propagator, 
with differences being the truncation of the integral 
and the use of an imaginary (Euclideanized) Schwinger proper time.
Transforming $\alpha$ into the real proper time $\tau$
corresponds to a complexification of 
the string worldsheet modular parameter~\cite{Witten:2013pra},
after which eq.~\eqref{stringy-Schwinger-1} becomes
\be
\mathcal{D}(k) 
= \int_{\ell_E^2}^{\infty} d\tau \, e^{- i \tau k^2} \, ,
\label{stringy-Schwinger-2}
\ee
where
\be
\ell_E^2 = - i\ell^2.
\label{ell-AC}
\ee
Notice that the exponential factor $e^{- i \tau k^2}$ does not lead to UV divergences
for both space-like and time-like momenta,
removing the need for Wick rotation.

We propose a prescription for analytic continuation 
in the light-cone quantization as follows: 
All integrals over the Lorentzian momentum space 
are initially evaluated with $\ell_E^2 > 0$, 
and subsequently, the obtained results are analytically continued 
as functions of $\ell_E^2$ onto the imaginary axis $\ell_E^2 = -i\ell^2$.
This approach preserves the Lorentzian signature of spacetime.

Based on this implementation,
the 2D stringy Feynman propagator~\eqref{G(X)} is thus
\begin{align}
G(U, V) 
&=
\int_{-\infty}^{\infty} \frac{d\Om d\bar{\Om}}{2\pi^2} \, 
\int_{\ell_E^2}^{\infty} d\tau \, e^{i 4 \tau\Om\bar{\Om}} \,
e^{- i\Om U - i\bar{\Om} V}
\nn \\
&=
\int_{-\infty}^{\infty} \frac{d\Om}{\pi} \, 
e^{- i\Om U}
\int_{\ell_E^2}^{\infty} d\tau \,
\delta(4\tau\Om - V) \, .
\end{align}
Ignoring ingoing modes with $\Om = 0$,
we arrive at
\begin{align}
G_{\text{out}}(U, V)
&=
\int_0^{|V|/4\ell_E^2} \frac{d\Om}{2\pi} \, 
\frac{1}{2\Om} 
\left[
\Th(V) \, e^{- i\Om U} 
+ \Th(- V) \, e^{i\Om U}
\right] \, .
\label{G-VE-2}
\end{align}
As expected from the findings in Ref.~\cite{Erler:2004hv},
the expression above manifests locality in one light-cone coordinate 
(in this case $V$)
but exhibits nonlocality in the other light-cone coordinate ($U$),
since the conjugate momentum $\Om$ of $U$ 
is subjected to a UV cutoff at $|V|/4\ell_E^2$.

Through eq.~\eqref{G=W+W}, 
we can now identify the stringy Wightman function as
\begin{align}
W_{\text{out}}(U, V)
= 
\int_{0}^{|V|/4\ell_E^2} \frac{d\Om}{2\pi} \, 
\frac{1}{2\Om} \, 
e^{- i\Om U}
\label{stringy-W-def}
\end{align}
for $V > 0$.
Imposing the reality condition on the field leads to
\be
W^{\ast}_{\text{out}}(x - x') 
= \langle 0 | \hat{\tilde{\phi}}_{\text{out}}(x) \hat{\tilde{\phi}}_{\text{out}}(x') | 0 \rangle^{\ast}
= \langle 0 | \hat{\tilde{\phi}}_{\text{out}}(x') \hat{\tilde{\phi}}_{\text{out}}(x) | 0 \rangle
= W_{\text{out}}(x' - x) \, ,
\ee
thus implying that eq.~\eqref{stringy-W-def} also applies when $V < 0$.
The stringy Wightman function differs from 
that of the low-energy effective theory~\eqref{W-EFT}
merely by the cutoff in the light-cone momentum $\Om$.
Given two spacetime points $(U, V)$ and $(U', V')$ 
separated by $(\Delta U, \Delta V) \equiv (U - U', V - V')$, 
eq.~\eqref{stringy-W-def} indicates that the outgoing modes 
contributing to the two-point correlation function $W_{\text{out}}(\Delta U, \Delta V)$
are those with momenta below $\Om_{\text{max}} = |\Delta V| / 4 \ell_E^2$. 
As this cutoff value is proportional to $|\Delta V|$, 
eq.~\eqref{stringy-W-def} demonstrates a UV/IR connection 
featured in the stringy model.
In particular, to probe the UV limit in the $U$-direction, 
the IR limit should be taken in the $V$-direction. 
We will explore this aspect further in section~\ref{sec:SpacetimeUncertaintyRelation}.


\subsection{Algebra of creation and annihilation operators}
\label{sec:FieldOperator}


While the calculation of Hawking radiation only requires 
knowledge of the Wightman function~\eqref{W=VEV},
the latter assumes the existence of a Hamiltonian formulation.\footnote{
The Hamiltonian formulation for nonlocal theories has 
been studied in various contexts (see, e.g., Refs.~\cite{Yang:1950vi, Umezawa-Takahashi, Hayashi,
Cheng:2001du, Cheng:2002rz, Llosa:1993sj, HerediaPimienta:2023ogb}).
However, to the best of our knowledge,
the Hamiltonian formulation for the stringy model~\eqref{L-def-2} as a quantum field theory
that is non-perturbative in $e^{-\ell^2 \Box}$
has never been found in the past.
}
To gain a deeper insight into the stringy model, 
we shall construct a consistent mode expansion 
for the field operator $\hat{\tilde{\phi}}_{\text{out}}$ within the framework of 
the stringy free-field theory, 
including the operator algebra of the associated creation and annihilation operators.

Let us derive the algebra of the creation and annihilation operators 
from the stringy Wightman function~\eqref{stringy-W-def},
which should be interpreted as
\begin{align}
W_{\text{out}}(\Delta U, \Delta V) 
= \langle 0 | \hat{\tilde{\phi}}_{\text{out}}(U, V) \hat{\tilde{\phi}}_{\text{out}}(U', V') | 0 \rangle
\label{W-VEV}
\end{align}
in the Hamiltonian formulation of the quantum stringy model.
Decomposing the field operator into Fourier modes as
\be
\hat{\tilde{\phi}}_{\text{out}}(U, V) = 
\int_0^{\infty} \frac{d\Om}{\sqrt{4\pi\Om}} 
\left(
a_{\Om}(V) \, e^{-i\Om U} + a^{\dag}_{\Om}(V) \, e^{i\Om U}
\right) \, ,
\label{phi-Om-VE-2}
\ee
eqs.~\eqref{stringy-W-def} and~\eqref{W-VEV} then result in
\begin{align}
\langle 0 | a_{\Om}(V) a^{\dag}_{\Om'}(V') | 0 \rangle
&= \frac{\sqrt{\Om\Om'}}{\pi} \int_{-\infty}^{\infty} dU dU' \,
e^{i \Om U} e^{- i \Om' U'} W_{\text{out}}(\Delta U, \Delta V)
\nn \\
&= \Th(|\Delta V| - 4 \ell_E^2 \Om) \, \delta(\Om - \Om') \, .
\label{aa-1}
\end{align}
Defining the vacuum state by
\be
a_{\Om}(V) | 0 \rangle = 0 \, ,
\label{vacuum-def}
\ee
we can evaluate $\langle 0 | \comm*{a_{\Om}(V)}{a_{\Om'}^{\dag}(V')} | 0 \rangle$,
and similarly the vacuum expectation values of other commutators.
We deduce that
\begin{align}
[a_{\Om}(V), a^{\dag}_{\Om'}(V')] &= \Th(|\Delta V| - 4 \ell_E^2\Om) \, \delta(\Om - \Om') \, ,
\label{CCR-aa-VE-1}
\\
[a_{\Om}(V), a_{\Om'}(V')] &= [a^{\dag}_{\Om}(V), a^{\dag}_{\Om'}(V')] = 0 \, .
\label{CCR-aa-VE-2}
\end{align}
These commutation relaions have the correct low-energy limit $\ell_E^2 \rightarrow 0$.

It may come as a surprise to the reader 
that the creation and annihilation operators are dependent on $V$.
It implies that the wave equation 
$e^{- \ell^2 \Box} \, \Box \hat{\tilde{\phi}} = 0$ 
derived from the action~\eqref{L-def-2} via the principle of least action
is not satisfied by the field operator $\hat{\tilde{\phi}}$.
This is a consequence of the nonlocal nature of the stringy model.
For the same reason,
the Wightman function~\eqref{stringy-W-def} 
does not satisfy the wave equation as well;
that is,
$e^{4 \ell^2 \del_U \del_V} \del_U\del_V W_{\text{out}}(U, V) \neq 0$.

An interesting implication of eqs.~\eqref{vacuum-def} and~\eqref{CCR-aa-VE-1} is that
the norm of the one-particle state $a^{\dag}_{\Om}(V) | 0 \rangle$ 
vanishes for any $\Om > 0$.
To define a state with a non-vanishing norm,
we need to superpose $a^{\dag}_{\Om}(V)$ at different values of $V$.
More generally,
a one-particle state is typically
a superposition of $\Om$-eigenstates:
\be
|\Psi \rangle \equiv 
\int d\Om \, \psi(\Om) |\Psi_{\Om} \rangle \, ,
\label{Psi-state}
\ee
where $|\Psi_{\Om} \rangle$ is taken to be $a_{\Om}^{\dag}|0\rangle$
in the low-energy effective theory.
However, 
since $a^{\dag}_{\Om}(V)$ is now $V$-dependent,
a generic $\Om$-eigenstate can have the form
\be
|\Psi_{\Om} \rangle \equiv
\int dV f_{\Om}(V) \, a_{\Om}^{\dag}(V) | 0 \rangle \, .
\label{state-ex}
\ee
It has a norm proportional to
\begin{align}
\langle \Psi_{\Om} | \Psi_{\Om} \rangle 
\propto \int dV dV' f_{\Om}^{\ast}(V) f_{\Om}(V') \, \Th(|\Delta V| - 4\ell^2_E\Om) \, ,
\label{norm-ex}
\end{align}
which vanishes if $f_{\Om}(V)$ has a finite support 
whose width is less than $4 \ell^2_E \Om$.
In other words, in the stringy model,
an outgoing mode with frequency $\Om$ 
must be defined over a range of $\abs{\Delta V} > 4 \ell_E^2 \Om$.
In contrast, in the low-energy effective theory,
an outgoing mode can be defined at an initial instant of time $V$
(with $\Delta V = 0$).


\subsection{Spacetime uncertainty principle}
\label{sec:SpacetimeUncertaintyRelation}


We established in section~\ref{sec:Introduction} that 
measuring the field $\phi$ is essentially a detection of $\tilde{\phi}$,
as all measurements rely on interactions.
Therefore, the Wightman function~\eqref{stringy-W-def} for $\tilde{\phi}$
reveals the need for a significant time interval $\abs{\Delta V}$
in order to detect a particle with a large frequency $\Omega$
in the stringy model.
We emphasize, though, that 
the discussions around eqs.~\eqref{Psi-state}--\eqref{norm-ex} suggest that
the mere existence of a non-zero norm state with a large $\Om$
already necessitates a substantial $\abs{\Delta V}$,
regardless of how or whether a measurement is carried out.
This constraint is in the form of a UV/IR connection:
In a region of Minkowski space restricted to 
a finite range $\abs{\Delta V}$ in the $V$-direction, 
the frequency $\Om$ of a quantum mode has a UV cutoff
\be
\Om \leq
\Om_{\text{max}}(\Delta V) \equiv \frac{\abs{\Delta V}}{4\ell_E^2} 
\label{Om<Ommax}
\ee
such that approaching the UV limit where $\Om \rightarrow \infty$
demands the IR limit $\abs{\Delta V} \rightarrow \infty$.

Since a UV cutoff in $\Om$ is equivalent to having a minimal uncertainty $\Delta U$ 
in the light-cone coordinate $U$,
eq.~\eqref{Om<Ommax} can be interpreted as giving rise to an uncertainty relation\,\footnote{
It may appear that the uncertainty relation~\eqref{DUDV} is potentially problematic
as we analytically continue $\ell_E^2$ to $-i\ell^2$ at the end of each calculation.
We will see in the next section that,
in the calculation of Hawking radiation,
there is only a negligible difference between writing $\ell_E^2$ or $\ell^2$
in the Wightman function \eqref{stringy-W-def}.
}
\be
\Delta U \Delta V \gtrsim 4\ell^2_E \, ,
\label{DUDV}
\ee
which is in the same form as the {\em spacetime uncertainty principle}
\be
\Delta T \Delta X \gtrsim \ell^2 
\label{DTDX}
\ee
proposed by Yoneya~\cite{Yoneya:1987gb, Yoneya:1988, Yoneya:1989ai, Yoneya:2000bt}.
It is shown in appendix~\ref{sec:STUR} that 
eq.~\eqref{DTDX} (with $\ell^2$ replaced by $2 \, \ell_E^2$) 
is also directly reflected in the stringy Feynman propagator~\eqref{G(X)}.

A physical interpretation of the spacetime uncertainty principle~\eqref{DTDX}~\cite{Yoneya:1987gb, Yoneya:1988, Yoneya:1989ai, Yoneya:2000bt} is that
high-energy strings are long, hence a large spatial uncertainty.
Similarly, eqs.~\eqref{Om<Ommax} and~\eqref{DUDV} indicate that
a substantial light-cone momentum $P_U = \Om$ 
can only be carried by a string with 
a large extension $\Delta V$ in the $V$-direction.
From an intuitive standpoint, one can imagine that 
for an outgoing mode with a large $\Om$, 
the background geometry is effectively probed by an extended string, 
thereby reducing its resolution.
This conceptual picture is jusified 
by an explicit calculation presented in appendix~\ref{sec:Background-Interaction},
where it is demonstrated perturbatively that 
the coupling between the field $\tilde{\phi}$ 
and a background fluctuation with a characteristic length scale $\lambda$ 
is suppressed when $\Delta V \geq 4 \ell_E^2 \Omega \gg \lambda$.
Therefore, 
eqs.~\eqref{Om<Ommax} and~\eqref{DUDV} showcase a pivotal aspect of the stringy model, 
namely that not only do long-wavelength modes perceive a smeared background, 
but this holds true for high-frequency modes as well.

As an extreme scenario (which will be relevant later), 
consider a quantum mode with an extraordinarily high frequency, 
to the extent that it requires to be defined on a scale 
as large as the entire universe.
In this case, 
the mode becomes insensitive to the precise configurations 
of individual black holes or galaxies.
A more appropriate approach for accurately describing 
the evolution of such a mode is to adopt 
the Friedmann-Robertson-Walker (FRW) metric 
to represent the background geometry, 
essentially rendering the black holes and galaxies akin to negligible specks of dust. 
In particular, 
this high-frequency quantum mode cannot be localized near the horizon of a black hole.
Rather, 
it is more accurate to state that it is defined 
in the asymptotically flat region of the black hole.
Hence, 
such a mode is not expected to contribute to Hawking radiation.
This insight turns out to have significant implications 
for the magnitude of late-time Hawking radiation, 
a subject we will delve into in detail in the next section.

Nevertheless, it is worth mentioning that, a priori, 
there could be an alternative interpretation of the stringy model, 
in which one considers $\phi$ as the fundamental field, 
with $\tilde{\phi}$ playing a role solely in interactions (see eq.~\eqref{L-def-1}).\footnote{
It is not clear whether this interpretation is realized
in a consistent Hamiltonian formulation,
but we cannot rule out this possibility at this stage.
}
In this perspective, 
an outgoing mode of $\phi$ can be defined at any given moment of $V$ (with $\Delta V = 0$), 
just as is the case in the low-energy effective theory, 
without the necessity of imposing any constraints on $\Delta V$.\footnote{
This is the case if we assume that there are 
creation operators other than $a_{\Om}^{\dag}(V)$
that can be used to define  
high-frequency one-particle states at an instant of $V$.
} 
The requirement for a bound on $\Delta V$ only arises 
when it comes to the detection of this outgoing mode.
As a result, 
the UV/IR relation~\eqref{Om<Ommax} applies exclusively to measurement processes,
and we would not be able to make the argument that Hawking radiation is turned off
around the scrambling time in the next section.

That said,
we should not take this viewpoint
if we do not believe in hidden degrees of freedom that can never be detected.
For example,
after imposing the periodic boundary condition of period $2\pi R$ on $V$,
$\Delta V$ is bounded from above by $2\pi R$.
As a result, 
only propagating modes with $\Om \leq \pi R/2\ell^2$ can be detected.
But, according to this alternative interpretation,
all modes with $\Om > \pi R/2\ell^2$ still exist in the theory.
On the other hand,
if we quantize $\tilde{\phi}$ as the fundamental field
and impose eqs.~\eqref{CCR-aa-VE-1} and~\eqref{CCR-aa-VE-2},
what is non-detectable is also non-existent.


\section{Hawking Radiation}
\label{sec:HawkingRadiation}


It has been previously argued in Ref.~\cite{Ho:2022gpg} that
the spacetime uncertainty principle~\eqref{DTDX} leads to
the termination of Hawking radiation around the scrambling time.
Roughly speaking,
Hawking radiation arises primarily due to the exponential blueshift,
a geometric effect that occurs as the Hawking modes 
are traced backwards in time close to the horizon.
A substantial uncertainty in space would 
effectively dilute the effects of this blueshift,
resulting in the suppression of Hawking radiation.
In this section, we will reach the same conclusion for the stringy model
through an explicit calculation, 
offering a more comprehensive explanation to 
substantiate the argument presented in Ref.~\cite{Ho:2022gpg}.


\subsection{Background geometry}
\label{sec:BackgroundGeometry}


In the conventional model of an evaporating black hole,
it is assumed that the quantum correction to the energy-momentum tensor
is sufficiently weak so that 
the Schwarzschild metric is a good approximation.\footnote{
For our purpose,
the back reaction on the Schwarzschild geometry is mild
assuming that the energy-momentum tensor is bounded from above by
$\langle T^{\mu}{}_{\mu} \rangle,
\langle T^{\th}{}_{\th} \rangle
\lesssim \mathcal{O}(1 / \ell_p^2 \, a^2)$~\cite{Ho:2019qiu}, 
where $\ell_p$ is the Planck length.
}
The four-dimensional Schwarzschild metric is 
\begin{align}
ds^2 &\equiv
- \left( 1 - \frac{a}{r} \right) dt^2 + \left( 1 - \frac{a}{r} \right)^{-1}  dr^2 + r^2 d\Omega_2^2
\nn \\
&= - \left( 1 - \frac{a}{r} \right) du \, dv + r^2 d\Omega_2^2 \, ,
\label{Schwarzschild-metric}
\end{align}
where $d\Omega_2^2 = d\th^2 + \sin^2\th \, d\varphi^2$ is the metric of a unit 2-sphere,
and the Eddington-Finkelstein light-cone coordinates $(u, v)$ are defined by
\begin{align}
u \equiv t - r_{\ast} \, , 
\qquad
v \equiv t + r_{\ast} \, ,
\label{uv-def}
\end{align}
with $r_{\ast}$ being the tortoise coordinate\,\footnote{
Our definition of the tortoise coordinate differs slightly from the usual one
by a shift of $a$ in order to simplify some equations that follow.
}
\be
r_{\ast}(r) \equiv
r - a + a \log\left(\frac{r}{a} - 1\right).
\label{rast-r}
\ee

In terms of the Kruskal light-cone coordinates defined by
\begin{align}
U(u) \equiv
- 2a e^{- u / 2a} \, ,
\qquad
V(v) \equiv
2a e^{v / 2a} \, ,
\label{U-u}
\end{align}
the Schwarzschild metric~\eqref{Schwarzschild-metric} is equivalent to
\begin{align}
ds^2 = - \frac{a}{r} \, e^{-(r-a)/a} \, dU dV + r^2 d\Omega_2^2 \, ,
\label{ds2-UV}
\end{align}
with the horizon located at $r = a$ (i.e. $U = 0$).
In the near-horizon region where $0 \leq r - a \ll a$,
the metric is approximately
\be
ds^2 \simeq - dU dV + r^2 d\Omega_2^2 \, .
\label{NHR-metric}
\ee
Upon the $s$-wave reduction,
this metric reduces to that of Minkowski space in two dimensions,
with the Kruskal coordinates playing the role of 
the light-cone coordinates in flat spacetime.

For simplicity,
we assume that the collapsing matter forming the black hole 
is a thin shell falling at the speed of light.
Without loss of generality,
we set the trajectory of the null shell as
\begin{align}
v = 0 \, ,
\quad
\mbox{or equivalently}
\quad 
V = 2a \, ,
\label{V-shell}
\end{align}
through a translation of the coordinate $t$.
Consequently,
$t = 0$ (or $u = 0$) corresponds to the moment when
the collapsing shell reaches $r_{\ast} = 0$,
where $r \simeq 1.567 a$.

The region inside the collapsing null shell is Minkowski space,
in which we can define the light-cone coordinates
\begin{align}
U = T - r \, , \quad V = T + r \, ,
\label{barU-barV}
\end{align}
where $T$ is the Minkowski time coordinate 
and $r$ is the radial coordinate.
This coordinate system inside the shell can be 
smoothly patched with the Kruskal coordinates 
in the near-horizon region outside the shell 
by shifting $T$ such that the shell intersects the horizon at $T = r = a$. 
That way, the horizon is situated at $U(T, r) = 0$, 
and the null shell coincides with $V(T, r) = 2a$, 
aligning with the setup~\eqref{V-shell} discussed earlier. 

So far, 
the stringy model has only been formulated in flat spacetime.
Therefore, 
before delving into the examination of Hawking radiation in this model,
it is crucial to note the range of the Kruskal coordinates $U$ and $V$ 
within which the background remains approximately flat. 
From the line element~\eqref{ds2-UV}, 
it is clear that this requires both $r - a \ll a$ as well as $r \simeq (V - U)/2$. 
According to eqs.~\eqref{uv-def}--\eqref{U-u}, 
the former condition is equivalent to 
\be 
- UV = (2a)^2 e^{r/a}\left(\frac{r - a}{a}\right) \ll (2a)^2
\qquad \text{for} \quad U < 0 \, ,
\label{cond-1}
\ee 
whereas the latter condition demands that 
\be
V \simeq 2r - |U| \qquad \text{for} \quad U < 0 \, .
\ee
The region relevant to the evolution of late-time (large $u$) Hawking quanta is $\abs{U} \ll a$.
In this case, the second condition on the areal radius 
further constrains that $V - 2a \ll a$. 
Hence, only a small neighborhood just outside the shell 
where $\abs{U} \ll a$ and $\Delta V \ll a$, 
along with the region inside the shell, 
can be treated as Minkowski space. 
The maximum range $\Delta V$ of $V$ near the horizon, 
beyond which the approximation loses its validity, 
is roughly
\be
\Delta V \sim \mathcal{O}(a) \, .
\ee


\subsection{Hawking radiation in low-energy effective theory}
\label{sec:DerivationHawkingRadiationEFT}


In this subsection,
we will revisit the derivation of Hawking radiation 
in the low-energy effective theory. 
This review will serve as a foundation for 
our subsequent calculation regarding the stringy model 
in the following subsection. 

Hawking radiation is characterized by 
the expectation value of the number operator 
\be
n_{\psi} \equiv b^{\dag}_{\psi} b_{\psi}
\label{npsi-def}
\ee
associated with Hawking particles described by a given wave function $\psi$.
For an outgoing Hawking particle
localized around the retarded time $u = u_0$
and with central frequency $\omega_0$, 
it can be represented by a wave function of the form
\be
\psi(u) \equiv \int d\om \, f^{\ast}_{\om_0}(\om) \, \frac{e^{- i\om(u - u_0)}}{\sqrt{4 \pi \omega}} \, ,
\label{psi-def}
\ee
where the profile function $f_{\om_0}(\om)$ is 
concentrated around $\om_0$ and has a width of $\Delta\om$ in the frequency domain.
The annihilation operator corresponding to the wave function $\psi$ 
is defined as
\begin{align}
b_{\psi}
&\equiv
\int d\om \, f_{\om_0}(\om) \, e^{- i\om u_0} \, b_{\om} \, ,
\label{bpsi-def}
\end{align}
where
\begin{align}
b_{\om}
&\equiv 
\sqrt{\frac{\om}{\pi}} \int du \, e^{i\om u} \, \phi(u ,v)
\qquad
(\om > 0) \, .
\label{b-def}
\end{align}
In the low-energy effective theory,
we do not distinguish $\phi$ from $\tilde{\phi}$. 
With $f_{\om_0}(\om)$ appropriately normalized according to
\be
\int_0^{\infty} d\om \, |f_{\om_0}(\om)|^2 = 1 \, ,
\label{f-normalization}
\ee
it follows that $[b_{\psi}, b^{\dag}_{\psi}] = 1$.

Although we have written $\phi = \phi(u ,v)$ in the definition~\eqref{b-def} of $b_{\omega}$, since the outgoing sector of the field is purely a function of $u$, $b_{\psi}$ does not depend on $v$.
Thus,
in principle,
one can imagine measuring the observable $n_{\psi}$
in the near-horizon region
or even inside the collapsing shell.
In this sense, 
the detection of Hawking radiation can be achieved
by coupling a detector to the field operator,
analogous to the Unruh effect.
The reason why Hawking radiation is typically associated with 
a detection performed at large distances 
is because the operator $n_{\psi}$ agrees with 
the notion of particle number in the asymptotically flat region.

To calculate $\langle 0 | n_{\psi} | 0 \rangle$,
we first derive
\begin{align}
\langle 0 | b^{\dag}_{\om} b_{\om'} | 0 \rangle
&=
\frac{\sqrt{\om\om'}}{\pi} \int du \, e^{- i\om u} \int du' \, e^{i\om' u'} \, 
\langle 0 | \phi(u, v) \phi(u', v') | 0 \rangle
\nn \\
&=
\frac{\sqrt{\om\om'}}{\pi} \int du \, e^{- i\om u} \int du' \, e^{i\om' u'} \, 
\langle 0 | \Phi(U(u), V(v)) \Phi(U(u'), V(v')) | 0 \rangle \, ,
\end{align}
where the Kruskal coordinates $U(u)$ and $V(v)$ are defined in eq.~\eqref{U-u},
and $\Phi(U(u), V(v)) = \phi(u, v)$ is 
the same field expressed in terms of $U$ and $V$.
Using the Wightman function~\eqref{W-EFT}
\begin{align}
\langle 0 | \Phi(U, V) \Phi(U', V') | 0 \rangle
=
\int_0^{\infty} \frac{d\Om}{2\pi} \, \frac{1}{2\Om} \, e^{- i \Om (U - U')}
\label{Wightman-EFT}
\end{align}
for outgoing modes in the low-energy effective theory,
we find that 
\begin{align}
\langle 0 | b^{\dag}_{\om} b_{\om'} | 0 \rangle
&=
\frac{\sqrt{\om\om'}}{\pi}
\int_0^{\infty} \frac{d\Om}{2\pi} \, \frac{1}{2\Om} \, 
\int du \, e^{- i\om u} \, e^{- i \Om U(u)} 
\int du' \, e^{i\om' u'} \, e^{i \Om U(u')}
\nn \\
&=
\int_0^{\infty} d\Om\, 
\b^{\ast}_{\om\Om} \, \b_{\om'\Om} \, ,
\label{VEV-bbom}
\end{align}
where the Bogoliubov coefficient is
\begin{align}
\b_{\om\Om}
&=
\frac{1}{2\pi} \sqrt{\frac{\om}{\Om}} \int_{-\infty}^{\infty} du \, e^{i\om u} \, e^{i\Om U(u)}
= 
\frac{a}{\pi} \sqrt{\frac{\om}{\Om}} \left(2a\Om\right)^{i2a\om} e^{-\pi a\om} \, \Gamma(-i2a\om) \, .
\label{Bogo-beta}
\end{align}

Substituting eq.~\eqref{VEV-bbom} back into 
the quantity of interest, $\langle 0 | n_{\psi} | 0 \rangle$, 
and employing the definition~\eqref{bpsi-def}, 
we obtain the expression
\begin{align}
\langle 0 | n_{\psi} | 0 \rangle
&= 
\int d\om \, d\om' \, f_{\om_0}^{\ast}(\om) f_{\om_0}(\om') \, e^{i (\om-\om') u_0}
\int_0^{\infty} d\Om \,
\b^{\ast}_{\om\Om} \, \b_{\om'\Om} \, .
\label{VEV-n-m}
\end{align}
Let us assume that the profile function $f_{\om_0}(\om)$
possesses a narrow width $\Delta \om \ll \omega_0 \sim 1/a$ such that
functions of $\om$ in the integrand can be approximated by their values at $\om_0$,
provided they change slowly with $\om$ over scales not shorter than $1/a$.
This allows us to further rewrite eq.~\eqref{VEV-n-m} as 
\be 
\langle 0 | n_{\psi} | 0 \rangle
\simeq 
\frac{a^2}{\pi^2} \, \omega_0 \, e^{-2 \pi a \omega_0} \, \abs{\Gamma(i 2a \omega_0)}^2
\int d \omega \, d \omega' \, f_{\omega_0}^*(\omega) f_{\omega_0}(\omega') \, e^{i (\omega - \omega') u_0}
\int_0^{\infty} \frac{d\Omega}{\Omega} \, (a \Omega)^{-i 2 a (\omega - \omega')} \, .
\label{VEV-n-m-2}
\ee 
Next,
we replace the integration over $\Om$ with an integration over $u$
via the change of variable
\be
\Om = \frac{1}{a} \, e^{u / 2a} \, .
\label{Om-u-cov}
\ee
As a result,
eq.~\eqref{VEV-n-m-2} becomes
\begin{align}
\label{n-LEET}
\langle 0 | n_{\psi} | 0 \rangle
&\simeq
\frac{1}{2\pi} \frac{1}{e^{4\pi a\om_0} - 1} 
\int d \omega \, d \omega' \, 
f_{\omega_0}^*(\omega) f_{\omega_0}(\omega') 
\int_{-\infty}^{\infty} du \, e^{-i (\omega - \omega') (u - u_0)} \\
&= 
\frac{1}{e^{4\pi a\om_0} - 1} 
\int d\omega \left| f_{\omega_0}(\omega) \right|^2 \\
&=
\frac{1}{e^{4\pi a\om_0} - 1} \, ,
\end{align}
where we have utilized the normalization condition~\eqref{f-normalization}.
Thus, 
we arrive at the conventional result of a stationary Hawking radiation,
featuring a Planck distribution at the Hawking temperature $T_H = 1/(4\pi a)$.

In the derivation above, 
we have neglected interactions in the low-energy effective theory.
In general,
renormalizable interactions result in only minor modifications to Hawking radiation~\cite{Leahy:1983vb, Frasca:2014gua}. 
However, non-renormalizable, higher-derivative interactions can lead to exponentially large corrections to Hawking radiation~\cite{Ho:2020cbf, Ho:2020cvn, Ho:2021sbi, Ho:2022gpg}.
Nevertheless, 
this effect can be safely disregarded in the stringy model, 
thanks to the exponential suppression of trans-Planckian interactions. 
For this reason, 
our subsequent discussion will center solely on the stringy free-field theory.


\subsection{Hawking radiation in stringy model}
\label{sec:HawkingRadiationinStringyModel}


In this subsection, 
we will derive the Hawking radiation within the context of the stringy model.
Contrasting with the derivation presented in the previous subsection, 
the Wightman function~\eqref{W-EFT} of the low-energy effective theory 
should be replaced by the stringy Wightman function~\eqref{stringy-W-def}. 
Additionally, due to the nonlocal nature of the stringy model,
particle detection can only be achieved over a certain time interval.

A particle detector that operates within a finite range ${\cal V}$ in time $v$
can be described by a switching function $s(v)$, 
which has the property that 
$s(v) \simeq 0$ for $v \notin {\cal V}$ 
and $s(v) \simeq 1$ for $v \in {\cal V}$.
The expectation value of the number of Hawking particles 
detected in the state $\psi$ is then
\be
\langle n_{\psi}({\cal V}) \rangle
\equiv
\frac{1}{\left(\int_{\cal V} dv'' s(v'')\right)^2} 
\int_{\cal V} dv \, s(v) \int_{\cal V} dv' \, s(v') \, 
\langle 0 | n_{\psi}(v, v') | 0 \rangle \, ,
\label{dVdVn}
\ee
where the number operator is
\begin{align}
n_{\psi}(v, v')
\equiv
b^{\dag}_{\psi}(v) b_{\psi}(v') \, .
\end{align}
Following the calculation procedure outlined in the preceding subsection, 
we now incorporate the $v$-dependence of the operators 
$\{ b_{\omega}, b_{\omega}^{\dagger} \}$ 
and utilize the stringy Wightman function $W_{\text{out}}(U, V)$~\eqref{stringy-W-def}. 
This leads to the expression
\begin{align}
\langle 0 | b^{\dag}_{\om}(v) b_{\om'}(v') | 0 \rangle 
&=
\frac{\sqrt{\om\om'}}{\pi} \int du \, e^{- i\om u} \int du' \, e^{i\om' u'} \, 
\langle 0 | \Phi(U(u), V(v)) \Phi(U(u'), V(v')) | 0 \rangle \nn \\
&=
\frac{\sqrt{\om\om'}}{\pi} \int du \, e^{- i\om u} \int du' \, e^{i\om' u'} \, 
W_{\text{out}}(\Delta U, \Delta V) \nn \\
&= 
\int_0^{\Delta V / 4 \ell_E^2} d\Om \, \b^{\ast}_{\om\Om} \, \b_{\om'\Om} \, ,
\label{stringy_bb}
\end{align}
where 
\be
\Delta U \equiv U(u) - U(u') \, ,
\qquad
\Delta V \equiv |V(v) - V(v')| \, .
\label{DV-def}
\ee
The Kruskal coordinate $U(u)$ is defined in eq.~\eqref{U-u},
and we will comment more on the function $V(v)$ in the next subsection.
The sole distinction between eq.~\eqref{stringy_bb}
and its counterpart~\eqref{VEV-bbom} in the low-energy effective theory
is the presence of a UV cutoff $\Om_{\text{max}} = \Delta V/4\ell_E^2$ 
in the integration over $\Om$.
Consequently, instead of eq.~\eqref{VEV-n-m}, 
we obtain
\begin{align}
\langle 0 | n_{\psi}(v, v') | 0 \rangle
&= 
\int d\om \, d\om' \, f_{\om_0}^{\ast}(\om) f_{\om_0}(\om') \, e^{i (\om-\om') u_0} \,
\langle 0 | b^{\dag}_{\om}(v) b_{\om'}(v') | 0 \rangle
\nn \\
&= 
\int d\om \, d\om' \, f_{\om_0}^{\ast}(\om) f_{\om_0}(\om') \, e^{i (\om-\om') u_0}
\int_0^{\Delta V/4\ell_E^2} d\Om \, \b^{\ast}_{\om\Om} \, \b_{\om'\Om} \, .
\label{VEV-n-0}
\end{align}
Due to the translation symmetry in $V$ in the near-horizon region, 
$\langle 0 | n_{\psi}(v, v') | 0 \rangle$ depends on $v$ and $v'$ 
only through $\Delta V$~\eqref{DV-def}, 
so we shall denote it as 
$\langle 0 | n_{\psi}(\Delta V) | 0 \rangle$ going forward.
The quantity $\langle n_{\psi}({\cal V}) \rangle$ defined in eq.~\eqref{dVdVn} 
then represents the number of Hawking particles 
detected in the near-horizon region.
If $\langle n_{\psi}({\cal V}) \rangle$ vanishes in this region, 
it would imply the absence of Hawking particles at large distances as well.

Finally, by taking the same steps in eqs.~\eqref{VEV-n-m-2}--\eqref{n-LEET} 
and adopting the same approximation scheme as in the previous subsection, 
we end up with
\begin{align}
&\langle 0 | n_{\psi}(\Delta V) | 0 \rangle \nn \\
&\simeq
\frac{a^2}{\pi^2} \, \omega_0 \, e^{-2 \pi a \omega_0} \, \abs{\Gamma(i 2a \omega_0)}^2 
\int d\om \, d\om' \, f_{\om_0}^{\ast}(\om) f_{\om_0}(\om') \, e^{i (\om-\om') u_0} 
\int_0^{\Delta V / 4\ell_E^2} \frac{d\Om}{\Om} 
\left( a\Om \right)^{-i2a(\om-\om')} \nn \\
&=
\frac{a}{\pi} \frac{1}{e^{4\pi a\om_0} - 1}
\int d\om \, d\om' \, f_{\om_0}^{\ast}(\om) f_{\om_0}(\om') \, e^{i (\om-\om') u_0}
\int_{-\infty}^{u_{\text{cut}}(\Delta V)} \frac{du}{2a} \,
e^{-i(\om-\om')(u \, + \, i\pi a)}
\nn \\
&\simeq
\frac{2 \omega_0}{e^{4\pi a\om_0} - 1}
\int_{-\infty}^{u_{\text{cut}}(\Delta V)} du \,
|\psi(u)|^2 \, ,
\label{n-VEV}
\end{align}
where we have reintroduced the wave function $\psi(u)$~\eqref{psi-def} 
of the Hawking particle into the expression.
The result closely resembles eq.~\eqref{n-LEET} 
but with the addition of a cutoff at
\begin{align}
u_{\text{cut}}(\Delta V)
&\equiv
2a \log\left(\frac{a\Delta V}{4\ell^2}\right) .
\label{ucut-def}
\end{align}
In the calculation above,
we performed the analytic continuation 
$\ell_E^2 \rightarrow \ell^2 = i \ell^2_E$~\eqref{ell-AC},
while also applying the change of variable $u \rightarrow u + i\pi a$.
This introduced an extra factor $e^{(\om - \om') \pi a}$,
which is approximately $1$ due to the assumption that 
$f_{\om_0}(\om)$ has a narrow width $\Delta \om \ll \omega_0$.
With only minor adjustments, 
this calculation aligns closely with the one presented in Ref.~\cite{Ho:2022gpg}, 
where it was demonstrated that a UV cutoff in $\Om$ results in the termination of 
Hawking radiation around the scrambling time.

According to eq.~\eqref{dVdVn}, a finite detection range ${\cal V}$ in the $V$-direction places an upper bound $\Delta V_{\text{max}}$ on $\Delta V$.
On the other hand, the expectation value~\eqref{n-VEV} vanishes when the wave packet $\psi(u)$~\eqref{psi-def}, which is centered at $u = u_0$ with a width of $\Delta u$, lies outside the range of integration $u \in \bigl( -\infty, u_{\text{cut}}(\Delta V) \bigr)$.
As a result, the number of Hawking particles $\langle n_{\psi}({\cal V}) \rangle$~\eqref{dVdVn} vanishes when
\be
u_0 - u_{\text{cut}}(\Delta V_{\text{max}}) \gg \Delta u \, .
\label{cond-u-1}
\ee
For typical Hawking radiation with dominant frequency $\om_0 \sim 1/a$,
we need $\Delta u \gg a$ 
to ensure precise resolution of the frequency
(i.e. $\Delta \om \ll \om_0$).
For instance,
for $\om_0 = 1/a$ and $\Delta \om = \om_0/100$,
we have $\Delta u = 100 a$.
Therefore, we conclude that
Hawking radiation can no longer be detected when
\be
u - u_{\text{cut}}(\Delta V_{\text{max}}) \gg \mathcal{O}(a) \, .
\label{u-ucut}
\ee
Put differently, 
this signifies that to detect Hawking radiation at a late time (large $u$),
a large $\Delta V_{\text{max}}$ (an extended duration of detection ${\cal V}$)
is necessary in order to define the quantity 
$\langle n_{\psi}({\cal V}) \rangle$~\eqref{dVdVn}.

Strictly speaking,
the applicability of the stringy model is confined 
to the near-horizon region where 
the 4D background geometry is approximately flat 
(see Sec.~\ref{sec:BackgroundGeometry}).
Hence, 
our derivation of Hawking radiation in the stringy model 
remains reliable only up to a detection range 
$\Delta V_{\text{max}} \sim \mathcal{O}(a)$, 
which corresponds to the detection of Hawking quanta 
before the scrambling time: $u \lesssim u_{\text{cut}}(a) \simeq 4a \log(a / \ell)$.
On the other hand, 
it is often assumed that,
even when the 4D geometry is curved,
the local Minkowski vacuum approximation holds as long as 
the 2D geometry (the radial part) is approximately flat
(i.e. when eq.~\eqref{cond-1} is valid).
Naively, this would imply that
the condition~\eqref{u-ucut} can be easily bypassed,
allowing for the successful detection of Hawking radiation at all times.
More explicitly,
suppose that the range of detection is set to be 
$\mathcal{V} \in (v_i = 0, v_f)$, 
then using eq.~\eqref{U-u}, we have
$u_{\text{cut}}(\Delta V_{\text{max}}) \simeq v_f + 2a \log(a^2 / \ell^2) > v_f$.
As a result, 
to detect Hawking radiation at time $u$, 
it would suffice to choose $v_f > u$, 
which is easily attainable.
However, 
there are additional subtleties in the stringy model arising from the UV/IR connection~\eqref{Om<Ommax}.
As we will elucidate below, 
not only do $\Delta V_{\text{max}}$ and $v_f$ 
become roughly linearly related 
when dealing with high-frequency modes 
that require a range $\Delta V_{\text{max}} \sim v_f \gg \mathcal{O}(a)$,
rendering it practically impossible to detect late-time Hawking quanta, 
but Hawking radiation itself ceases to exist after the scrambling time.


\subsection{Hawking radiation after scrambling time}
\label{sec:HR-scramblingtime}


In the conventional description of Hawking radiation 
in the low-energy effective theory, 
Hawking quanta detected after the scrambling time 
originate from fluctuations with trans-Planckian frequencies 
$\Om \gg \mathcal{O}(a / \ell^2)$ in the Unruh vacuum 
(i.e. the local Minkowski vacuum) 
confined within a Planckian distance from the horizon.
In the case of the stringy model, 
the UV/IR connection~\eqref{Om<Ommax} between 
the frequency $\Omega$ and the associated minimal scale $\Delta V$ 
applies not only to detection processes 
but also to the very existence and definition of quantum modes, 
as elaborated in Sec.~\ref{sec:FieldOperator}. 
As a consequence, 
if there were to be late-time Hawking quanta, 
their trans-Planckian ancestors in the distant past 
would have spanned an interval 
$\Delta V \gg \mathcal{O}(a)$ in the $V$-direction.\footnote{
For a solar-mass black hole ($a \simeq 3$ km), 
a Hawking particle later than merely $u \simeq 5$ ms 
would have had a central frequency in the remote past 
corresponding to a $\Delta V$ that is 
already larger than the size of the universe! 
To put this into perspective, 
the conventional estimate of the black hole's lifetime is 
$\mathcal{O}(a^3 / \ell^2) \sim 10^{64}$ years.
}

\begin{figure}[t]
\centering
\includegraphics[scale=0.5]{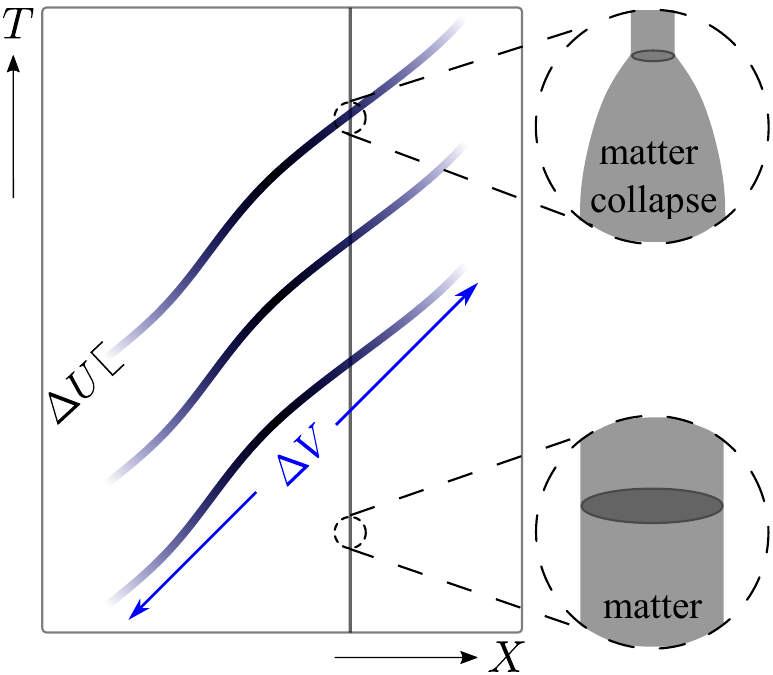}
\caption{
An outgoing wave packet with a width of $\Delta U$ 
can only be defined over a range of $\Delta V > 4\ell^2/\Delta U$.
The vertical line represents a region considerably smaller than $\Delta V$,
consisting of matter with low density that eventually collapses into a black hole.
Such a wave packet cannot be localized in the near-horizon region of the black hole,
and thus will not contribute to Hawking radiation
}
\label{fig:uv-ir}
\end{figure}

To illustrate this further,
suppose that a Hawking particle 
described by a wave packet with central frequency $\om \sim 1/a$ is 
detected at a retarded time $u = 2 (n + 1) a \log(a / \ell)$.
Let us trace this wave packet backward in time 
to the Minkowski region inside the shell,
and then following it all the way to the Minkowski spacetime in the far past.
During this backward evolution, 
the wave packet would have experienced a significant blueshift 
in the near-horizon region such that 
its central frequency $\Omega$ in the distant past would have been
\be
\Om(u) = \om \left(\frac{dU}{du}\right)^{-1} 
\sim \frac{1}{a} \, e^{u / 2a} \, .
\label{Om-om}
\ee
However, according to eq.~\eqref{Om<Ommax}, 
such a high-frequency quantum mode demands to be defined at the scale 
\be
\Delta V(u) 
\geq 4\ell^2\Om(u) 
\sim \frac{\ell^2}{a} \, e^{u / 2a}
\sim \left(\frac{a}{\ell}\right)^{n-1} a \, .
\label{deltaV_u}
\ee
For a macroscopic black hole ($a \gg \ell$),
this length scale vastly exceeds the size $a$ of the black hole for $n > 1$.
Consequently, in the Minkowski space of the far past, 
the black hole appears to be no more than a tiny speck of dust to this mode. 
This implies that the black hole background 
would have had little to no impact on the evolution of the mode in the first place, 
as illustrated in Fig.~\ref{fig:uv-ir}.
Given this, 
it becomes unlikely that this mode would eventually give rise to Hawking radiation.

In fact, for a solar-mass black hole, 
$\Delta V$~\eqref{deltaV_u} is of the scale of the observable universe 
for a value of $n$ as small as $1.26$!
It is shown in Appendix~\ref{sec:Background-Interaction} 
(and was also argued in Sec.~\ref{sec:SpacetimeUncertaintyRelation}) 
that the effects of background fluctuations 
with length scales much shorter than $\Delta V \geq 4\ell^2 \Om$
are suppressed for high-frequency modes $\Om$.
Therefore, on such an enormous scale,
all geometric structures, including black holes and galaxies, 
can be effectively treated as if they have been smoothed out. 
It is then natural to define the vacuum state of the high-frequency modes 
on the Friedmann-Robertson-Walker (FRW) background,
which is better approximated by the vacuum 
in the asymptotically flat region of a black hole,
as opposed to that in the near-horizon region.
Since this vacuum state is insensitive to the 
time-dependent background of black hole formation 
that occurs on a minuscule scale compared to the extent of these modes, 
it remains practically unchanged both in the far past and the far future.
As a result,
these modes cannot contribute to the particle creation process, 
and we conclude that Hawking radiation no longer exists 
after the scrambling time when $u \gtrsim 4a \log(a/\ell)$.\footnote{
For instance, 
we would anticipate Hawking radiation from a solar-mass black hole 
to be terminated around the time $4.5 a\log(a/\ell) \simeq 4$ ms. 
In fact, the existence of Hawking radiation in this scenario becomes questionable.
}

Last but not least,
let us briefly address the function $V(v)$ 
used in the derivation of Hawking radiation in Sec.~\ref{sec:HawkingRadiationinStringyModel}.
In principle, this function should correspond to the Kruskal coordinate $V(v)$ as defined in eq.~\eqref{U-u} for the Schwarzschild metric.
However, since the geometry is probe-dependent,
and high-frequency modes experience a smoother background,
the exponential relation between $V$ and $v$ would be effectively smeared to an approximately linear relation for high-frequency modes defined on immense scales $\Delta V \gg a$.


\section{Conclusion and Discussion}
\label{sec:conclusion}


Using a class of non-local quantum field theories,
we have analyzed how Hawking radiation is affected by
the stringy effect characterized by 
the exponential suppression of interactions in the ultraviolet regime.
Owing to the UV/IR relation~\eqref{Om<Ommax} 
constraining the frequency and spatial extent of quantum modes, 
the behavior of the trans-Planckian ancestors of 
late-time Hawking radiation differs significantly from 
that in the standard low-energy effective theory, 
where high-frequency modes are confined 
in the vicinity of the black hole horizon.
We argued based on this distinctive feature that 
the specific stringy effect under consideration 
results in the termination of Hawking radiation 
beyond the scrambling time.

In our setup, 
we have assumed a negligible back reaction of the vacuum energy-momentum tensor,
as in the conventional model of black hole evaporation.
Thus, our findings may not apply to models with 
a large vacuum energy-momentum tensor around the horizon
(e.g. the KMY model~\cite{Kawai:2013mda, Kawai:2014afa}).
Additionally, it is worth noting that our conclusion 
could potentially be influenced by stringy effects 
other than the suppression of high-energy interactions. 
Nonetheless, it suffices to say that 
the conventional model of black hole evaporation is, 
if not entirely ruled out, 
far from being confirmed by string theory.

As Hawking radiation lasts for only approximately the scrambling time,
the mass of the black hole decreases by merely a fraction of order
$\mathcal{O}\bigl( (\ell/a)^2 \log(a/\ell) \bigr)$ relative to its initial mass.
Consequently, 
the black hole remains essentially classical 
with just small quantum corrections,
and the information loss paradox is absent in this scenario.

Our conclusion stands in contrast to some common beliefs regarding Hawking radiation.
It is often asserted that the ``nice-slice'' argument~\cite{Polchinski:1995ta} 
suggests that Hawking radiation persists as an adiabatic process 
well described by the low-energy effective theory 
until the black hole becomes microscopic.
However, there exists a loophole in this argument~\cite{Ho:2021sbi}: 
the time evolution of a quantum mode with a trans-Planckian center-of-mass energy 
in a scattering process with the background curvature depends on UV physics, 
even when high-energy excitations are not present.
In the context of the stringy model discussed in this work,
even though the high-energy interactions are suppressed, 
the nice-slice argument is still inapplicable 
due to nonlocality inherent in the stringy model.

It was pointed out in Ref.~\cite{Blamart:2023ixr} that
an effective UV cutoff that shuts down Hawking radiation 
around the scrambling time is in accordance with 
the Trans-Planckian Censorship Conjecture (TCC)~\cite{Bedroya:2019snp} 
in the context of inflationary cosmology.
For reliable predictions about 
primordial perturbations on an inflationary background,
the phase of accelerated expansion 
should not last beyond the TCC time scale, 
analogous to the scrambling time for black holes.
Otherwise, the fluctuations we observe today 
would have been smaller than the Planck length 
in the early stages of inflation, 
rendering the low-energy effective theory invalid~\cite{Martin:2000xs}.
Similarly, 
if Hawking radiation originates from trans-Planckian modes, 
it cannot be reliably described within the framework of effective field theory. 
One way to circumvent this issue is by 
restricting Hawking radiation to occur within the scrambling time.

Notice that the final result~\eqref{n-VEV} of Hawking radiation 
presented in this work still displays a Planck spectrum 
characterized by the Hawking temperature $T_H = (4 \pi a )^{-1}$,
despite the decrease in the magnitude of the radiation over time.
The relation $dS_{BH} = (T_H)^{-1} dM$ then yields 
the same Bekenstein-Hawking entropy $S_{BH}$~\cite{Bekenstein:1973ur}, 
thereby respecting the universality of black hole thermodynamics~\cite{Mathur:2023uoe}. 
Our result is also not in direct contradiction to
recent developments in the computation of black hole entanglement entropy
using the prescription of quantum extremal surfaces~\cite{Penington:2019npb, Almheiri:2019psf, Almheiri:2019hni}.
This prescription allows us to calculate the entanglement entropy of Hawking radiation,
but is unable to unveil the ultimate fate of a specific initial state, 
which is a dynamical issue.
In this work, we have highlighted 
a potential source of nonlocality required by the notion of the {\em islands}.

Our investigation has revealed that 
the stringy model leads to 
a Lorentz-invariant uncertainty relation~\eqref{DUDV} with a minimum length scale,
which is the light-cone version of the spacetime uncertainty principle~\eqref{DTDX}~\cite{Yoneya:1987gb, Yoneya:1988, Yoneya:1989ai, Yoneya:2000bt}.
This is the essential reason leading to the peculiar result that
Hawking radiation terminates around the scrambling time.
However,
strictly speaking, 
this UV/IR connection, as well as other peculiar features,
could have their origins in certain pathologies of the model.
Our claims remain to be fully justified.
Nevertheless, we believe that this hardly discussed alternative resolution 
to the information loss paradox deserves more exploration.


\section*{Acknowledgements}

We thank Emil Akhmedov, Chi-Ming Chang, Yuhsi Chang, Ronny Chau, 
Henry Liao, Nobuyoshi Ohta, Cheng-Tsung Wang, and Tamiaki Yoneya
for valuable discussions. 
P.M.H. and W.H.S. are supported in part by the Ministry of Science and Technology, R.O.C.
(MOST 110-2112-M-002-016-MY3),
and by National Taiwan University. 
Y.I. was partially supported by Grand-in-Aid for Scientific Research (C) (No.\ 21K03569),
Ministry of Education, Science and Culture, Japan.
H.K. thanks Prof. Shin-Nan Yang and his family
for their kind support through the Chin-Yu chair professorship.
H.K. is partially supported by Japan Society of Promotion of Science (JSPS),
Grants-in-Aid for Scientific Research (KAKENHI)
Grants No.\ 20K03970 and 18H03708,
by the Ministry of Science and Technology, R.O.C. (MOST 111-2811-M-002-016),
and by National Taiwan University.


\appendix



\section{From Propagator to Spacetime Uncertainty}
\label{sec:STUR}


In this appendix,
we present evidence of Yoneya's spacetime uncertainty principle~\eqref{DTDX} in the stringy model.
We start with the Fourier transform of the stringy propagator~\eqref{propagator-Mink}
of a massless scalar $\tilde{\phi}$ with respect to the Minkowski time coordinate $T$:
\begin{align}
G_{\vb{K}}(T - T') 
&\equiv 
\int_{-\infty}^{\infty} dk^0 \,
e^{i k^0 (T - T')} \, {\cal D}(k) \\
\begin{split}
&=
\frac{\pi}{2} \,
\frac{1}{\left| \vb{K} \right|} \, 
\Bigg\{
e^{- i \left| \vb{K} \right| |T - T'|} \,
\mbox{erfc} \left[ (-1)^{1/4} \left( \left| \vb{K} \right| \ell_E  
- \frac{|T - T'|}{2 \, \ell_E} \right) \right] \\
&\qquad \qquad \quad 
+ \, e^{i \left| \vb{K} \right| |T - T'|} \,
\mbox{erfc} \left[ (-1)^{1/4} 
\left(\left| \vb{K} \right| \ell_E + \frac{|T - T'|}{2 \, \ell_E} \right) \right]
\Bigg\} \, .
\label{Fourier-Propagator}
\end{split}
\end{align}
Making use of the asymptotic behavior
\be
\left|\mbox{erfc}[(-1)^{1/4} z]\right| \simeq 2 \Th(- z)
\qquad \mbox{for} \quad z \in \mathbb{R} \quad \text{and} \quad |z| \gg 1 \, ,
\ee
we see that eq.~\eqref{Fourier-Propagator} agrees with the low-energy effective theory
in the limit $\ell_E \rightarrow 0$,
and that it approximately vanishes when
\be
\left| \vb{K} \right| \ell_E
- \frac{|T - T'|}{2 \, \ell_E} \gg 1 \, .
\label{STUR-origin}
\ee
That is, 
the propagator~\eqref{Fourier-Propagator} vanishes unless
\be
|\vb{K}| \lesssim \frac{\Delta T}{2 \, \ell^2_E} \, ,
\label{k_max}
\ee
where $\Delta T \equiv |T - T'|$.
Since an upper bound on the magnitude $|\vb{K}|$ of the spatial momentum
is equivalent to a lower bound on the uncertainty $\Delta X$ in position
defined in the direction of $\vb{K}$,
eq.~\eqref{k_max} leads to the spacetime uncertainty relation~\cite{Yoneya:1987gb, Yoneya:1988, Yoneya:1989ai, Yoneya:2000bt}
\be
\Delta T \Delta X \gtrsim 2 \, \ell^2_E \, .
\label{STUR-1}
\ee


\section{Interaction With the Background}
\label{sec:Background-Interaction}


Let us consider a background interaction term
\begin{align}
S_I &= 
\frac{1}{2} \int d^2 x \, h(V) \, \tilde{\phi}^2(U, V)
\nn \\
&= 
\frac{1}{2} \int dV \, h(V) \int_0^{\infty} \frac{d\Om}{2\Om} \,
a_{\Om}(V) a^{\dag}_{\Om}(V) \, ,
\label{Lint-def-3}
\end{align}
where $h(V)$ is a background field assumed to be constant in $U$ for simplicity,
and we have used the mode expansion~\eqref{phi-Om-VE-2} of $\tilde{\phi}$.
The total action for the field $\tilde{\phi}$ is now $S = S_0 + S_I$,
where $S_0$ is the free-field action for the stringy model:
\begin{align}
S_0 &=
\frac{1}{2} 
\int d^2 x \, \tilde{\phi} \, \Box \, e^{-\ell^2 \Box} \, \tilde{\phi} \nn \\
&= 
i \int dV \int_0^{\infty} d\Omega \, a_{\Omega}^{\dagger}(V - 4 \ell_E^2 \Omega) \, \del_V a_{\Omega}(V) \, ,
\end{align}
with the analytic continuation $\ell^2 \to \ell_E^2 = -i \ell^2$ performed.

As an illustrative example, 
we examine the impact of the interaction~\eqref{Lint-def-3} 
between the field $\tilde{\phi}$ and a background fluctuation described by
\be
\label{h}
h(V) = h_0 \, e^{i \lambda^{-1} V} 
\ee
with a small amplitude $h_0$.
In the lowest-order approximation with respect to $h_0$,
we obtain
\begin{align}
&\langle 0 | a_{\Om}(V) a^{\dag}_{\Om'}(V') | 0 \rangle \nn \\
\begin{split}
&\simeq 
\epsilon( \abs{V - V'} - 4\ell_E^2 \Om) \, \d(\Om - \Om') \\
&\quad 
+ \frac{i}{4 \Om} \, h_0 \, \d(\Om - \Om')
\int_{-\infty}^{\infty} dV'' \, \eps(V'') \,
\Th(- \abs{V - V'} + 8 \ell_E^2 \Om - V'') \\
&\qquad \qquad \qquad \qquad \qquad \quad 
\times 
\left[ \Theta(V - V') \, e^{i \lambda^{-1} (V'' + V - 4 \ell_E^2 \Omega)} - \Theta(V' - V) \, e^{-i \lambda^{-1} (V'' + V' - 4 \ell_E^2 \Omega)} \right]
\, ,
\label{h_corr}
\end{split}
\end{align}
where $\epsilon(z) \equiv \Theta(z) - 1/2$.
The question is whether the second term in this expression, 
arising from the background interaction, 
becomes effectively suppressed at large $\Om$ when $\ell_E \neq 0$.

In a physical scenario, 
we typically have a wave packet of the field $\tilde{\phi}$ 
characterized by a finite width $\Delta \Omega$ in frequency 
on a background consisting of a superposition of fluctuations~\eqref{h} 
with a finite width $\Delta \lambda$ in wavelength. 
The presence of the oscillatory factors 
$e^{\pm i 4 \lambda^{-1} \ell_E^2 \Om}$ in eq.~\eqref{h_corr} then
suppresses the effect of the background field on $\tilde{\phi}$
when $4 (\Delta \lambda)^{-1} \ell_E^2 \Om \gg 1$ 
or when $4 \lambda^{-1} \ell_E^2 (\Delta \Om) \gg 1$.
For instance, on a Schwarzschild background, 
where the characteristic length scale is the Schwarzschild radius, 
we have $\lambda \sim \Delta \lambda \sim \mathcal{O}(a)$ 
for the background fluctuations.
Consequently, 
the coupling of the field to the Schwarzschild background 
becomes significantly suppressed when $\Omega \gg a/ 4 \ell_E^2$,
or equivalently, when $\abs{\Delta V} \geq 4\ell_E^2 \Om \gg a$, 
as indicated by the UV/IR relation~\eqref{Om<Ommax}.
This is the case for Hawking particles, 
as their corresponding wave packets 
not only exhibit blueshifted central frequencies $\Omega$ in the past 
but also encompass a broad spectrum $\Delta\Omega \gtrsim \Omega$~\cite{Ho:2021sbi}. 
Therefore, the deviation from flat spacetime caused by 
the black hole background is highly suppressed for Hawking modes.


\end{document}